\documentclass[envcountsect,envcountsame]{llncs}
\pdfoutput=1
\usepackage[normal]{optional}

\usepackage{amsmath}
\usepackage{amssymb}
\usepackage{commath}
\usepackage{wrapfig}
\usepackage[noadjust]{cite}
\usepackage{graphicx}

\newcommand{\marginwidth}{1.15in}
\usepackage[margin=\marginwidth,centering]{geometry}
\usepackage[normalem]{ulem}

\opt{sub}{\newcommand{\figureSize}{\scriptsize}}
\opt{normal}{\newcommand{\figureSize}{\footnotesize}}

\newcommand{\rethm}[2]{\paragraph{{\bf Theorem #1.}}{\emph{#2}}}
\newcommand{\relem}[2]{\paragraph{{\bf Lemma #1.}}{\emph{#2}}}

\renewcommand{\norm}[1]{\enVert[0]{#1}}

\newenvironment{Proof}{\begin{proof}}{\qed\end{proof}}

\newcommand{\pre}{\mathsf{pre}}
\newcommand{\post}{\mathsf{post}}
\newcommand{\N}{\mathbb{N}}
\newcommand{\Z}{\mathbb{Z}}
\renewcommand{\emptyset}{\varnothing}

\newcommand{\rest}[1]{\hspace{-1mm}\restriction_{#1}}

\newcommand{\calN}{\mathcal{N}}
\newcommand{\calD}{\mathcal{D}}

\newcommand{\calI}{\mathcal{I}}
\newcommand{\calO}{\mathcal{O}}
\newcommand{\calL}{\mathcal{L}}
\newcommand{\calV}{\mathcal{V}}
\newcommand{\calM}{\mathcal{M}}

\opt{normal}{\renewcommand{\vec}[1]{{\bf #1}}}

\newcommand{\vb}{{\vec{b}}}
\newcommand{\vi}{{\vec{i}}}
\newcommand{\vc}{{\vec{c}}}
\newcommand{\vv}{{\vec{v}}}
\newcommand{\vr}{{\vec{r}}}
\newcommand{\vp}{{\vec{p}}}
\newcommand{\vo}{{\vec{o}}}
\newcommand{\vx}{{\vec{x}}}
\newcommand{\vy}{{\vec{y}}}

\spnewtheorem{observation}[theorem]{Observation}{\bfseries}{\itshape}

\newcommand{\reachone}{\Rightarrow}
\newcommand{\reach}{\Rightarrow^*}

\title{Democratic, Existential, and Consensus-Based Output Conventions in Stable Computation by Chemical Reaction Networks\thanks{The first author is a postdoctoral fellow of the Research Foundation -- Flanders (FWO). The second author was supported by NSF grant CCF-1619343, and the third author by NSF grants 
CCF-1618895 and CCF-1652824.}}
\titlerunning{Output Conventions in Stable Computation by Chemical Reaction Networks}
\author{Robert Brijder\inst{1} \and David Doty\inst{2} \and David Soloveichik\inst{3}}
\institute{Hasselt University, Diepenbeek, Belgium \email{robert.brijder@uhasselt.be} 
\and University of California, Davis, CA, USA \email{doty@ucdavis.edu} 
\and University of Texas, Austin, TX, USA \email{david.soloveichik@utexas.edu}}
\begin{document}

\maketitle

\begin{abstract}
We show that some natural output conventions for error-free computation in chemical reaction networks (CRN) lead to a common level of computational expressivity.
Our main results are that the standard consensus-based output convention have equivalent computational power to
(1) \emph{existence-based}
and
(2) \emph{democracy-based}
output conventions.
The CRNs using the former output convention have only ``yes'' voters, with the interpretation that the CRN's output is yes if any voters are present and no otherwise.
The CRNs using the latter output convention define output by majority vote among ``yes'' and ``no'' voters.

Both results are proven via a generalized framework that simultaneously captures several definitions,
directly inspired by a Petri net result of Esparza, Ganty, Leroux, and Majumder [CONCUR 2015].
These results support the thesis that the computational expressivity of error-free CRNs is intrinsic, not sensitive to arbitrary definitional choices.
\end{abstract}

\section{Introduction}
Turing machines solve exactly the same class of yes/no decision problems whether they report output via accept/reject states, or if instead they write a 1 or 0 on a worktape before halting.
Similarly, finite-state transducers compute the same class of functions whether they emit output
on a state (\emph{Moore machine}~\cite{moore1956gedanken})
or a transition (\emph{Mealy machine}~\cite{mealy1955method}).
In general, if the power of a model of computation is insensitive to minor changes in the definition,
this lends evidence to the claim that the model is robust enough to apply to many real situations,
and that theorems proven in the model reflect fundamental truths about reality,
rather than being artifacts of arbitrary definitional choices.

The theory of chemical reaction networks (CRNs) studies the general behavior of chemical reactions in well-mixed solutions, abstracting away spatial properties of the molecules.
Formally, a CRN is defined as a finite set of reactions such as $2A+C \to 2B$, where $A$, $B$, and $C$ are abstract chemical species.
In a discrete CRN the state of the system is given by molecule counts of each species and the system updates by application of individual reactions.

CRNs have only recently been considered as a model of computation \cite{DBLP:journals/nc/SoloveichikCWB08},
motivated partially by the ability to implement them using a basic experimental technique called \emph{DNA strand displacement}~\cite{SimCRN/Soloveichik}.
Discrete CRNs with standard stochastic kinetics are Turing complete if allowed an arbitrary small, but nonzero, probability of error~\cite{DBLP:journals/nc/SoloveichikCWB08},
improved to error probability 0 in \cite{DBLP:conf/dna/CummingsDS14}.\footnote{We always assume that the given CRN reactions are obeyed perfectly; even so if reactions happen to occur in a certain inauspicious order, an incorrect output might be obtained.
It is beyond the scope of this paper to consider imperfect physical realizations of CRNs, in which spurious reactions outside of the desired CRN can occur (see e.g.~\cite{robust-pp}).
}
It is known that an error-free computational model of CRNs inspired by the theory of population protocols~\cite{dblp:journals/dc/angluinadfp06,dblp:conf/podc/angluinae06} decides exactly the semilinear sets (that do not contain the zero vector)~\cite{DBLP:journals/dc/AngluinAER07}.\footnote{When the set of configurations reachable from an initial configuration is always finite (for instance, with population protocols, or more generally mass-conserving CRNs), then error-freeness coincides with error probability 0. See \cite{DBLP:conf/dna/CummingsDS14} for an in-depth discussion of how these notions can diverge when the set of configurations reachable from an initial configuration is infinite.}

We study the computational robustness of error-free CRNs under different output conventions.
The original output convention~\cite{dblp:journals/dc/angluinadfp06} for deciding predicates (0/1-valued functions)
is that each species is classified as voting either 0 (``no'') or 1 (``yes''),
and a configuration (vector of nonnegative integer counts of each species) $\vo$ has output $i\in\{0,1\}$ if all species present in positive count are $i$-voters, i.e., there is a \emph{consensus} on vote $i$.
As an example, the CRN with reactions $X_1 + N \to Y$ and $X_2 + Y \to N$, with initial configuration $\{x_1 X_1,x_2 X_2,1 N\}$,
where $N,X_2$ vote 0 and $Y,X_1$ vote 1, decides if $x_1 > x_2$; $Y$ and $N$ alternate being present as each reacts with an input, so the first input to run out determines whether we stop at $Y$ or $N$.
More formally, we say $\vo$ is \emph{output-stable} if every configuration $\vo'$ reachable from $\vo$ has the same output as $\vo$ (i.e., the system need not halt, but it stops changing its output).
Finally, it is required that a correct output-stable configuration is reachable not only from the initial configuration $\vi$, but also from any configuration reachable from $\vi$; under mild assumptions (e.g., conservation of mass), this implies that a correct stable configuration is actually reached with probability 1 under the standard stochastic kinetic model~\cite{Gillespie77}.
It has been shown in \cite{dblp:journals/dc/angluinadfp06} that the computational power is not reduced, that is, it still decides precisely all semilinear sets, when we restrict to those CRNs where
(1) each reaction has two reactants and two products (e.g., disallowing reactions such as $2A+C \to 2B$ and $A \to B+C$, a model known as a \emph{population protocol}~\cite{dblp:journals/dc/angluinadfp06})
and
(2) the system eventually halts for every possible input (see also \cite{DBLP:conf/dna/Brijder14}).

One can imagine alternative output conventions, i.e., ways to interpret what is the output of a configuration, while retaining the requirement that a correct output-stable configuration is reachable from any reachable configuration.
Rather than requiring every species to vote 0 or 1,
for example,
allow the CRN to designate some species as nonvoters.
It is not difficult to show
\opt{normal}{(see Section~\ref{sec:sym-nonvoters})}
that such CRNs have equivalent computational power:
They are at least as powerful since one can always choose all species to be voters.
The reverse direction follows by converting a CRN with a subset of voting species into one in which every species votes, by replacing every nonvoting species $S$ with two variants $S_0$ and $S_1$,
whose voting bit is swayed by reactions with the original voting species, and which are otherwise both functionally equivalent to $S$.

We investigate two output conventions that are not so easily seen to be convertible to the original convention.
The first convention is \emph{existence-based}, in which there are only 1-voters, whose presence or absence indicates a configuration-wide output of 1 or 0, respectively.
It is not obvious how to convert such an existential CRN into a consensus-based CRN, since this appears to require producing 0-voters if and only if 1-voters are absent.
The second convention is \emph{democracy-based}, in which there are 0- and 1-voters, but the output of a configuration is given by the majority vote rather than being defined only with consensus.
Intuitively, the difficulty in converting such a democratic CRN into a  consensus-based CRN is that,
although the democratic CRN may stabilize on a majority of, for example, 1-voters over 0-voters,
the exact numerical gap between them may never stabilize.
A straightforward attempt to convert a democratic CRN into a consensus CRN results in a CRN that changes the output every time a new 0- or 1-voter appears.
For instance, suppose we use the previously described CRN for computing whether $x_1 > x_0$,
where $x_1$ and $x_0$ respectively represent the count of 1- and 0-voters.
If the original democratic CRN repeatedly increments $x_0$ and then $x_1$,
the resulting CRN flips between $Y$ and $N$ indefinitely
--- thus never stabilizing in the consensus model ---
even if $x_1 > x_0$ remains true indefinitely.

We show that these conventions have equivalent power as the original definition.
Our techniques further establish that the class of predicates computable by CRNs is robust to two additional relaxations of the classical notion of stable computation~\cite{dblp:journals/dc/angluinadfp06}:
(1) a correct output configuration need not be reachable from \emph{every} reachable configuration, only the initial configuration, and
(2) the set of output configurations need not be ``stable'' (i.e., closed under application of reactions),
so long as each initial configuration can reach only a correct output.

After defining existing notions of computation by CRNs in Section~\ref{sec:crns_crds}, we introduce in Section~\ref{sec:gcrd} a very general computational model for CRNs, called a \emph{generalized chemical reaction decider} (gen-CRD).
Its definition is directly inspired by a recent powerful result from Petri net theory~\cite{DBLP:conf/concur/EsparzaGLM15,Esparza2016},
restated here as Theorem~\ref{thm:concur_petri_result}.
Using this result we show that under mild conditions, gen-CRDs decide only semilinear sets.
We then show that the original consensus-based model,
the existence-based model,
and the democracy-based model
all fit into this framework, establishing their common expressivity.

One reason to consider the democracy-based output convention is due to its propitious composition properties.
Analogous to wiring up pre-built circuit-boards in electronics, we would like to be able to create larger chemical computation by composing two pre-existing CRN modules.
Note that in the strand displacement implementation, mixing together two solutions implementing two different CRNs amounts to concatenating the CRNs: i.e., a new CRN that is the union of the chemical reactions of the two.
The problem is that given two error-free CRNs, such that the output species of one are the input species of the other,
it is not in general meaningful to concatenate them.
Intuitively there are two issues: (1) the downstream CRN may consume the output of the upstream CRN before the upstream CRN finishes, and interfere with the upstream computation;
(2) the upstream CRN may change the output before it stabilizes, but the downstream CRN may use the previous incorrect answer. 
Both problems can be avoided if the upstream CRN never consumes its output species~\cite{CheDotSolNaCo}.
For boolean inputs/outputs, avoiding consuming output species naturally leads to the democracy-based output convention, where the 0/1 value can be changed by producing more of the opposite output.

A conference version of this paper was presented at DNA~22 \cite{DNA22/BDS/2016}.

\section{Chemical reaction networks and deciders} \label{sec:crns_crds}

\subsection{Chemical reaction networks} \label{ssec:crns}

Let $\Z$ and $\N$ denote the integers and nonnegative integers, respectively.
Let $\Lambda$ be a finite set.
The set of vectors over $\N$ indexed by $\Lambda$ (i.e., the set of functions $\vc: \Lambda \rightarrow \N$) is denoted by $\N^\Lambda$.
The zero vector is denoted $\vec{0}$.
For $\vc,\vc' \in \N^\Lambda$ we write $\vc \leq \vc'$ if and only if $\vc(S) \leq \vc'(S)$ for all $S \in \Lambda$.
For $\vc \in \N^\Lambda$ and $\Sigma \subseteq \Lambda$, the \emph{projection} of $\vc$ to $\Sigma$, denoted by $\vc\rest{\Sigma}$, is an element in $\N^\Sigma$ such that $\vc\rest{\Sigma}(S) = \vc(S)$ for all $S \in \Sigma$.
Let $\norm{ \vc } = \norm{ \vc }_1 = \sum_{S \in \Lambda} \vc(S)$ denote the $L_1$ norm of $\vc$.
We sometimes use multiset notation, e.g., $\vc=\{1A,2C\}$ to denote $\vc(A)=1,\vc(C)=2,\vc(S)=0$ for $S \in \Lambda \setminus \{A,C\}$, or when defining reactions, additive notation, i.e., $A+2C$.

A \emph{reaction} $\alpha$ over $\Lambda$ is an ordered pair $(\vr,\vp)$ with $\vr,\vp \in \N^\Lambda$, where $\vr$ and $\vp$ are the \emph{reactants} and \emph{products} of $\alpha$, respectively.
We write $\vr \to \vp$ to denote a reaction $(\vr,\vp)$, e.g., $A+B \to 2A+C$ denotes the reaction $(\{A,B\},\{2A,C\})$.

\begin{definition}
A \emph{chemical reaction network (CRN)}  is an ordered pair $\calN = (\Lambda, R)$ with $\Lambda$ a finite set and $R$ a finite set of reactions over $\Lambda$.
\end{definition}
The elements of $\Lambda$ are called the \emph{species} of $\calN$.
The elements of $\N^\Lambda$ are called the \emph{configurations} of $\calN$.
Viewing $\vc$ as a multiset, each element of $\vc$ is called a \emph{molecule}.
For $\vc,\vc' \in \N^\Lambda$, we write $\vc \reachone_{\calN} \vc'$ if there is a reaction $\alpha = (\vr,\vp) \in R$ such that $\vr \leq \vc$ and $\vc' = \vc-\vr+\vp$.
The transitive and reflexive closure of $\reachone_{\calN}$ is denoted by $\reach_{\calN}$.
If $\calN$ is clear from the context, then we simply write $\reachone$ and $\reach$ for $\reachone_{\calN}$ and $\reach_{\calN}$, respectively.
If $\vc \reach \vc'$, then we say $\vc'$ is \emph{reachable} from $\vc$.

For $\vc \in \N^\Lambda$, we define $\pre_{\calN}(\vc) = \{ \vc' \in \N^\Lambda \mid \vc' \reach_\calN \vc \}$ and $\post_{\calN}(\vc) = \{ \vc' \in \N^\Lambda \mid \vc \reach_\calN \vc' \}$.
Again we omit the subscript $\calN$ if the CRN $\calN$ is clear from the context.
Note that for $\vc,\vc' \in \N^\Lambda$, we have $\vc \in \pre(\vc')$ if and only if $\vc' \in \post(\vc)$ if and only if $\vc \reach \vc'$.
We extend $\pre(\vc)$ and $\post(\vc)$ to sets $X \subseteq \N^\Lambda$ in the natural way:
$\pre(X) = \bigcup_{\vc \in X} \pre(\vc)$ and $\post(X) = \bigcup_{\vc \in X} \post(\vc)$.

Petri net theory is a very well established theory of concurrent computation \cite{petrinet/review/pet1977}.
We recall here that CRNs are essentially equivalent to Petri nets.
In Petri net terminology, molecules are called ``tokens'', species are called ``places'', reactions are called ``transitions'', and configurations are called ``markings''. Due to this correspondence, we can apply results from Petri net theory to CRNs (which we will do in this paper, cf.\ Theorem~\ref{thm:concur_petri_result}). Conversely, the results shown in this paper can be reformulated  straightforwardly in terms of Petri nets.
Vector addition systems~\cite{VASsKarpMiller} form a model nearly equivalent to CRNs and Petri nets, where reactions roughly correspond to vectors with integer entries.\footnote{The only difference is \emph{catalysts}: reactants that are also products, e.g., $C+X \to C+Y$, are allowed in CRNs and Petri nets but not in vector addition systems. Most results for these models are insensitive to this difference.}
In the special case of population protocols~\cite{dblp:journals/dc/angluinadfp06},
each reaction $\alpha = (\vr,\vp)$ obeys $\norm{\vr}=\norm{\vp}=2$.
As a result, for each configuration $\vc$ of a population protocol, both $\pre(\vc)$ and $\post(\vc)$ are finite (because there are only a finite number of configurations $\vc'$ with $\norm{\vc'} = \norm{\vc}$).
In that model, molecules are called ``agents'', species are called ``states'', and reactions are called ``transitions''.

\subsection{Consensus-based output-stable deciders} \label{ssec:crds}

We now recall how one can compute using CRNs. Say we want to decide whether or not the number $n$ of molecules of species $X$ is even. One way to do this is by introducing the reaction $X+X \to \emptyset$.\footnote{Notation $\emptyset$ indicates that this reaction has no products.} If $n$ is even, then eventually all molecules are consumed, and if $n$ is odd, then eventually there is exactly one molecule of species $X$ present.
Once the CRN has stabilized, the presence of a molecule of species $X$ signals that $n$ is odd (i.e., there were an odd number of molecules of species $X$ present initially). Note that in this example there is no molecule of any species that signals that $n$ is even. One may think of a more elaborate example where the presence of say, a molecule of species $V_{\mathrm{even}}$, signals (once the CRN has stabilized) that $n$ is even. In this way, once the CRN has stabilized, $X$ ``votes'' that $n$ is odd, while $V_{\mathrm{even}}$ ``votes'' that $n$ is even.

A chemical reaction decider $\calD$ (introduced in \cite{CheDotSolNaCo}) is a reformulation in terms of CRNs of the notion of population protocol~\cite{dblp:journals/dc/angluinadfp06} from the field of distributed computing.
We define a set of input configurations $\calI$ and two sets of ``trap configurations'',
called \emph{output-stable} configurations, $\calO_0$ and $\calO_1$.
We then say that $\calD$ is \emph{output-stable} and \emph{decides} the set $\calI_1 \subseteq \calI$ (with $\calI_0 = \calI \setminus \calI_1$) if for each $i \in \{0,1\}$
(1)
starting from a configuration in $\calI_i$, the CRN remains always within reach of a configuration in $\calO_i$ (i.e., $\post(\calI_i) \subseteq \pre(\calO_i)$),
and (2)
once a configuration is in $\calO_i$, it is stuck in $\calO_i$ (i.e., $\post(\calO_i) = \calO_i$).

The sets $\calI$, $\calO_0$, and $\calO_1$ are all of a specific form.
There is a subset of \emph{input} species $\Sigma \subseteq \Lambda$; $\calI$ consists of nonzero configurations where the all molecules present are in $\Sigma$.
The output is based on consensus: all the molecules present in an output configuration must agree on the output.
More precisely, there is a partition $\{\Gamma_0,\Gamma_1\}$ of $\Lambda$ (called \emph{0-voters} and \emph{1-voters}, respectively),%
\footnote{The definition of~\cite{CheDotSolNaCo} allows only a subset of $\Lambda$ to be voters, i.e., $\Gamma_0 \cup \Gamma_1 \subseteq \Lambda$.
This convention is more easily shown to define equivalent computational power than our main results about existential and democratic voting.
\opt{normal}{See Section~\ref{sec:sym-nonvoters} for details.}%
}
such that configuration $\vc$ has \emph{output} $i \in \{0,1\}$ if all molecules present in $\vc$ are from $\Gamma_i$ (i.e., $\vc\rest{\Gamma_{1-i}}=\vec{0}$) and $\vc \neq \vec{0}$).
A configuration $\vo$ is defined to be in $\calO_i$ --- it is \emph{output-stable} --- if all configurations of $\post(\vo)$ also have output $i$.

Our definition, though equivalent,
is phrased differently from the usual one~\cite{dblp:journals/dc/angluinadfp06},
being defined in terms of $\calI$, $\calO_0$, and $\calO_1$ instead of $\Sigma$, $\Gamma_0$, and $\Gamma_1$.
This simplifies our generalization of this notion in Section~\ref{sec:gcrd}.

\begin{definition}\label{def:os_crd}
A \emph{consensus-based output-stable chemical reaction decider (con-CRD)} is a $4$-tuple $\calD = (\calN,\calI,\calO_0,\calO_1)$, where $\calN=(\Lambda,R)$ is a CRN and there are $\Sigma \subseteq \Lambda$ and a partition $\{\Gamma_0,\Gamma_1\}$ of $\Lambda$ such that
    \begin{enumerate}
    \item \label{defn:o-stable-crd-I}
    $\calI = \{ \vc \in \N^\Lambda \mid \vc\rest{\Lambda\setminus\Sigma} = \vec{0} \} \setminus \{\vec{0}\}$,

    \item \label{defn:o-stable-crd-O}
    $\calO_i = \{ \vc \in \N^\Lambda \mid \post(\vc) \subseteq \calL_i \setminus \calL_{1-i} \}$,
    with
    $\calL_i = \{ \vc \in \N^\Lambda \mid \vc\rest{\Gamma_i} \neq \vec{0} \}$ for $i \in \{0,1\}$.

    \item \label{defn:o-stable-crd-preI-in-postO}
    There is a partition $\{\calI_0,\calI_1\}$ of $\calI$ such that $\post(\calI_i) \subseteq \pre(\calO_i)$ for $i \in \{0,1\}$.

    \end{enumerate}
\end{definition}

\newcommand{\figSymCRDWidth}{0.7\textwidth}
\begin{figure}[ht]
    \centering
    \includegraphics[width=\figSymCRDWidth]{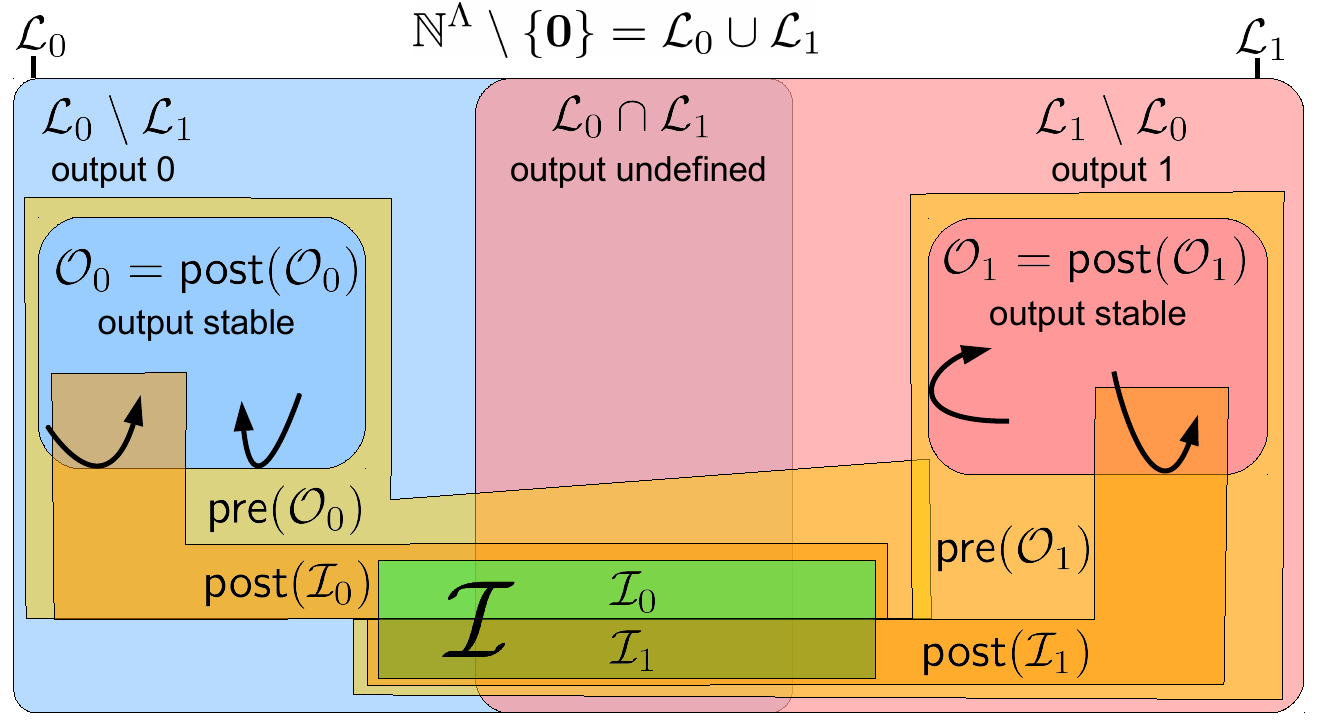}
    \caption{\figureSize Venn diagram of configurations that define con-CRD.
    Subset relationships depicted in their most general form:
    $\calI_i \subseteq \post(\calI_i) \subseteq \pre(\calO_i)$,
    and $\calO_i \subseteq \calL_i \setminus \calL_{1-i}$.
    $\pre(\calO_0)$ and $\pre(\calO_1)$ partition the set $\calI = \calI_0 \cup \calI_1$. The arrows, depicting possible trajectories in the set of configurations when reactions take place, illustrate that, once inside $\mathcal{O}_i$, we cannot escape from this set. 
    }
    \label{fig:crd-defn}
\end{figure}

Condition~\ref{defn:o-stable-crd-I} states that only species in $\Sigma$ may be present initially, and at least one must be present.
Condition~\ref{defn:o-stable-crd-O} defines $\calL_i$ to be configurations with an $i$-voter, so those in $\calL_i \setminus \calL_{1-i}$ unanimously vote $i$, and those in $\calO_i$ are stable (``stuck'' in the set $\calL_i \setminus \calL_{1-i}$).
Condition~\ref{defn:o-stable-crd-preI-in-postO} states that from every configuration reachable from an initial configuration, a ``correct'' output-stable configuration is reachable from there; this is the usual way of expressing stable computation~\cite{DBLP:journals/dc/AngluinAER07,CheDotSolNaCo}.
The relationships between these sets are illustrated in Figure~\ref{fig:crd-defn}.

\begin{remark}
A different definition is found in~\cite{CheDotSolNaCo} and a number of other papers.
That definition relaxes ours in two ways:
(1) having both voting and non-voting species,
(2) allowing non-input species in the input configuration (e.g., $\{1N\}$ in the Introduction).
\opt{normal}{In Appendix~\ref{sec:sym-nonvoters}, we show that (1) does not affect the computational power of the model.}%
\opt{sub}{It turns out that (1) does not affect the computational power of the model.}
It is also known~\cite{dblp:journals/dc/angluinadfp06} that (2) does not alter the computational power (though it may affect the time complexity~\cite{angluin2008fast,LeaderElectionDISC}).
\end{remark}

\begin{remark} \label{rem:Oi-pre-Lotheri}
We can equivalently define $\calO_i = \N^\Lambda \setminus \pre(\calL_{1-i} \cup \{\vec{0}\})$,
a form that will be useful later.
To see that this definition is equivalent, observe that $\N^\Lambda \setminus \calO_i$ is the set of configurations from which it is possible either to reach $\calL_{1-i}$, or to reach \emph{outside} of $\calL_i$,
and the only point outside \emph{both} is $\vec{0}$, so $\N^\Lambda \setminus \calO_i = \pre(\calL_{1-i} \cup \{\vec{0}\})$.
Thus $\calO_i = \N^\Lambda \setminus \pre(\calL_{1-i} \cup \{\vec{0}\})$.
\end{remark}

\begin{remark}
The $\calO_i$ are disjoint and closed under application of reactions: $\calO_0 \cap \calO_1 = \emptyset$ and $\post(\calO_i) = \calO_i$.
\end{remark}

\begin{remark}  \label{rem:weaker-calI}
Definition \ref{def:os_crd} implies the (weaker) condition that
$\calI_i = \calI \cap \pre(\calO_i)$.
This can be shown as follows.
First, $\calI_i \subseteq \calI$ and $\calI_i \subseteq \post(\calI_i) \subseteq \pre(\calO_i)$, so $\calI_i \subseteq \calI \cap \pre(\calO_i)$.
To see the reverse containment, assume $\vc \in \calI \cap \pre(\calO_i)$, but $\vc \notin \calI_i$,
i.e., $\vc \in \calI_{1-i} \cap \pre(\calO_i)$.
Let $\vo \in \post(\vc)$ be such that $\vo \in \calO_i$; such $\vo$ exists since $\vc \in \pre(\calO_i)$.
Since $\vo \in \post(\calI_{1-i}) \subseteq \pre(\calO_{1-i})$, we have $\vo \in \calO_{i} \cap \pre(\calO_{1-i})$.
Let $\vo'\in\post(\vo)$ such that $\vo'\in\calO_{1-i}$.
Then $\vo' \in \post(\calO_{i}) \cap \calO_{1-i}$ --- a contradiction because $\post(\calO_{i}) = \calO_{i}$ is disjoint from $\calO_{1-i}$.
\end{remark}

Since $\calI_0 = \calI \cap \pre(\calO_0)$ and $\calI_1 = \calI \cap \pre(\calO_1)$ are disjoint,
we say that a con-CRD $\calD$ \emph{decides} the set $\calI_1$.
If a con-CRD $\calD$ decides the set $X \subseteq \N^\Lambda$, then the entries indexed by $\Lambda \setminus \Sigma$ are zero for each $\vc \in X$.
Therefore, by abuse of notation, we also say that $\calD$ \emph{decides} the set $X\rest{\Sigma} \subseteq \N^\Sigma$.
We will use this convention for all chemical reaction deciders with $\calI$ of the given form.

\begin{example}\label{ex:symCRD}
We construct a con-CRD $\calD$ that decides the set $x \not\equiv y \mod m$ where $x$ and $y$ are non-negative integer variables, not both zero, and $m \geq 2$ is an integer constant.
The variables $x$ and $y$ represent initial counts of species $X$ and $Y$, respectively.
Let $\Sigma = \{X,Y\}$, $\Gamma_0 = \{V_0\}$, $\Gamma_1 = \{X,Y\}$, and $\Lambda = \Gamma_0 \cup \Gamma_1$ be as in Definition~\ref{def:os_crd}, with the following reactions:
\begin{eqnarray}
mX \to V_0, \quad mY \to V_0, \quad X+Y \to V_0, \label{eqn:simplify} \\
Y + V_0 \to Y, \quad X + V_0 \to X \label{eqn:make_false}.
\end{eqnarray}
We argue that $\calD$ decides the set $\{ \vc \in \mathbb{N}^\Sigma \setminus\{\vec{0}\} \mid \vc(X) \not\equiv \vc(Y) \mod m \}$.
Indeed, if $x \equiv y \mod m$, then eventually all $X$ and $Y$ molecules are consumed by the reactions of (\ref{eqn:simplify}).
The last time one of these reactions occurs introduces a $V_0$ molecule (there is a last reaction since $x$ and $y$ are not both zero).
So eventually we obtain a configuration $\vc \in \calL_0 \setminus \calL_1$ for which no reaction can be applied anymore.
Thus $\vc \in \calO_0$.
If $x \not\equiv y \mod m$, then eventually we reach a configuration with one of $X$ or $Y$, but not both, remaining.
The remaining $X$ or $Y$ molecules consume all $V_0$ molecules by the reactions of (\ref{eqn:make_false}),
without the possibility of producing any more.
So eventually we obtain a configuration $\vc' \in \calL_1 \setminus \calL_0$ for which no reaction can be applied anymore.
Thus $\vc' \in \calO_1$.
\end{example}

\subsection{Semilinear sets} \label{ssec:semilinear}

We say that $X\subseteq \N^\Lambda$ is \emph{linear} if there is a finite set
$\{\vv_1,\ldots,\vv_k\} \subseteq \N^\Lambda$ and $\vb \in \N^\Lambda$ such that
$X = \{ \vb + \sum_{i=1}^k n_i \vv_i \mid n_1,\ldots,n_k \in \N \}$.
We say that $X\subseteq \N^\Lambda$ is \emph{semilinear} if $X$ is the union of a finite number of linear sets.
Semilinear sets are precisely the sets definable in Presburger arithmetic, which is the first-order theory of natural numbers with addition.
As a consequence, the class of semilinear sets is closed under union, intersection, complementation, and projection~\cite{ginsburg1966}.
A useful characterization of semilinear sets is that they are exactly the sets expressible as finite unions, intersections, and complements of sets of one of the following two forms:
\emph{threshold sets} of the form $\{ \vx \in \N^\Lambda \mid \sum_{i \in \Lambda} a_i \cdot \vx(i) < b \}$
for some constants $a_i \in \Z$, with $i \in \Lambda$,
or
\emph{mod sets} of the form $\{ \vx  \in \N^\Lambda \mid \sum_{i \in \Lambda} a_i \cdot \vx(i) \equiv b \mod c \}$
for some constants $a_i \in \Z$, with $i \in \Lambda$, and $b,c\in\N$.

The following result was shown in \cite{dblp:journals/dc/angluinadfp06,dblp:conf/podc/angluinae06}. In fact, the result was shown for output-stable population protocols, which form a subclass of the con-CRDs. However, the proof is sufficiently general to hold for con-CRDs as well.\footnote{%
Indeed, the negative result of~\cite{dblp:conf/podc/angluinae06} that con-CRDs decide only semilinear sets is more general than stated in Theorem~\ref{thm:ostable_semilinear}, applying to \emph{any} reachability relation $\reach$ on $\N^\Lambda$ that is
reflexive, transitive, and ``additive'' ($\vx \reach \vy$ implies $\vx + \vc \reach \vy + \vc$).
Also, the negative result of~\cite{dblp:conf/podc/angluinae06} implicitly assumes that the zero vector $\vec{0}$ is not reachable (i.e., $\pre(\vec{0}) = \{\vec{0}\}$).
This assumption is manifest for population protocols (if the population size is non-zero).
For CRNs, this assumption can be readily removed;
see Lemma~\ref{lem:no_empty_prod}.
}

\begin{theorem}[\cite{dblp:journals/dc/angluinadfp06,dblp:conf/podc/angluinae06}] \label{thm:ostable_semilinear}
Let $X \subseteq \N^\Sigma \setminus \{\vec{0}\}$. Then $X$ is semilinear if and only if there is a con-CRD that decides $X$.
\end{theorem}

For a configuration $\vc \in \N^\Sigma$, $\pre(\vc)$ and $\post(\vc)$ are in general \emph{not} semilinear~\cite{DBLP:journals/tcs/HopcroftP79}.
Hence the semilinearity of Theorem~\ref{thm:ostable_semilinear} is due to additional ``computational structure'' of a con-CRD.
We repeatedly use the following notion of upwards closure to prove that certain sets are semilinear.
The results below were shown or implicit in earlier papers~\cite{DicksonsLemma,dblp:conf/podc/angluinae06}.
We say $X \subseteq \N^\Lambda$ is \emph{closed upwards} if, for all $\vc \in X$, $\vc' \geq \vc$ implies $\vc' \in X$.

\newcommand{\lemClosedUpwardsSemilinear}{Every closed upwards set $X \subseteq \N^\Lambda$ is semilinear.}
\newcommand{\proofLemClosedUpwardsSemilinear}{
    \begin{Proof}
    For each $\vb \in \min(X)$ we consider the linear set $L_\vb  = \{ \vb + \sum_{i=1}^{|\Lambda|} n_i \vv_i \mid n_1,\ldots,n_{|\Lambda|} \in \N \}$ where the $\vv_i$'s are the $|\Lambda|$ unit vectors of $\N^\Lambda$.
    Now, $X = \bigcup_{\vb \in \min(X)} L_\vb$. Since $\min(X)$ is finite by Lemma~\ref{lem:dicksons_lemma}, $X$ is semilinear.
    \end{Proof}
}
\newcommand{\lemClosedUpwardsPrePost}{If $X \subseteq \N^\Lambda$ is closed upwards, then so are $\pre(X)$ and $\post(X)$.}
\newcommand{\proofLemClosedUpwardsPrePost}{
    \begin{Proof}
    Let $\vc \in \pre(X)$ and $\vc' \geq \vc$.
    We show that $\vc' \in \pre(X)$.
    Let $\vec{d} = \vc' - \vc$.
    Since $\vc \in \pre(X)$, there exists $\vc''$ such that $\vc \reach \vc''$ and $\vc'' \in X$.
    Thus $\vc' = \vc + \vec{d} \reach \vc'' + \vec{d}$.
    Since $X$ is closed upwards, $\vc'' + \vec{d} \in X$, so $\vc' \in \pre(X)$.
    The $\post(X)$ case is symmetric.
    \end{Proof}
}

\newcommand{\dicksonsLemma}{
    For $X \subseteq \N^{\Lambda}$, define $\min(X) = \{ \vc \in X \mid (\forall \vc' \in X) \ \vc' \leq \vc \implies \vc' = \vc\}$ to be the \emph{minimal elements} of $X$.

    \begin{lemma}[Dickson's lemma \cite{DicksonsLemma}] \label{lem:dicksons_lemma}
    For all $X \subseteq \N^{\Lambda}$, $\min(X)$ is finite.
    \end{lemma}
}

\opt{normal}{\dicksonsLemma}

    \begin{lemma} \label{lem:closed_upw_semil}
    \lemClosedUpwardsSemilinear
    \end{lemma}

\opt{normal}{\proofLemClosedUpwardsSemilinear}

    \begin{lemma} \label{lem:closed_upw_pre_post}
    \lemClosedUpwardsPrePost
    \end{lemma}

\opt{normal}{\proofLemClosedUpwardsPrePost}

\newcommand{\argumentPreZeroNotSemilinear}{%
$\pre(\vec{0})$ is not semilinear for every CRN.
Hopcroft and Pansiot~\cite{DBLP:journals/tcs/HopcroftP79} show that $\post(\vc)$ may be non-semilinear:
they define $\vc=\{1P,1Y\}$ and reactions $P+Y \to P+X$, $P \to Q$, $Q+X \to Q+2Y$, $Q \to P+A$, with
$\post(\vc) = \{\vc \mid 0 < \vc(X)+\vc(Y) \leq 2^{\vc(A)} \text{ or } 0 < 2\vc(X)+\vc(Y) \leq 2^{\vc(A)+1} \}$,
which is not semilinear.
To see that $\post(\vec{0})$ can be non-semilinear, modify this CRN by adding a fifth reaction $\emptyset \to P+Y$,
which applied to $\vec{0}$ reaches $\vc=\{1P,1Y\}$.
Moreover, the set $S = \{\vx \mid \vx(P)+\vx(Q)=1\}$ is semilinear, so if $\post(\vec{0})$ were semilinear, $S \cap \post(\vec{0})$ would be as well.
Since a second execution of $\emptyset \to P+Y$ permanently exits $S$,
we have that $S \cap \post(\vec{0}) = \post(\vc)$, i.e., non-semilinear.
By replacing all reactions with their reverse, we obtain a CRN such that $\pre(\vec{0})$ is not semilinear.
}

Our results require $\pre(\vec{0})$ to be semilinear.\opt{normal}{\footnote{\argumentPreZeroNotSemilinear}}
Observe that $\pre(\vec{0}) = \{\vec{0}\}$ if and only if for each reaction $\alpha = (\vr,\vp)$, $\vp = \vec{0}$ implies $\vr = \vec{0}$.
The next lemma
shows that we can assume this holds for con-CRDs without loss of generality.

\newcommand{\LemNoEmptyProd}{
For every con-CRD $\calD$, there is a con-CRD $\calD'$ deciding the same set such that, for each reaction $\alpha = (\vr,\vp)$ of $\calD'$, $\vp \neq \vec{0}$.
}
\begin{lemma}\label{lem:no_empty_prod}
  \LemNoEmptyProd
\end{lemma}

\newcommand{\proofLemNoEmptyProd}{
    Let $\calD$ be a con-CRD that decides a set $X$.
    Add to $\calD$ two new species $D_0$ and $D_1$.
    Species $D_i$ will function as a ``dummy'' $i$-voter.
    Remove the useless reaction $\vec{0} \to \vec{0}$ if it exists in $\calD$.
    Replace each reaction $\alpha: \vr \to \vec{0}$, where $\vr$ contains only $i$-voters, by $\alpha': \vr \to D_i$. Replace each reaction $\alpha: \vr \to \vec{0}$, where $\vr$ contains both $0$ and $1$-voters, by $\alpha': \vr \to D_0$.
    (The choice for $D_0$ here instead of $D_1$ is arbitrary.)
    Moreover, for every species $S$ we add the reactions $S+D_0 \to S$ and $S+D_1 \to S$. Let $\calD'$ be the obtained system.

    We see that $\calD$ and $\calD'$ operate similarly.
    The only difference is that in the latter $D_i$'s may be produced and consumed.
    Now, in $\calD$, once a configuration $\vo \in \calO_i$ is reached,
    we have that for each $\vo' \in \post_\calD(\vo)$,
    every molecule of $\vo'$ is an $i$-voter (this holds in particular for the case $\vo' = \vo$).
    A corresponding configuration $\vec{d}$ in $\calD'$ may have some additional dummy molecules of species $D_{1-i}$.
    But eventually, these molecules will all be removed by the reactions $S+D_{1-i} \to S$.
    So, it suffices to verify that no $D_{1-i}$ molecule may be \emph{produced} in some $\vec{d}' \in \post_{\calD'}(\vec{d})$.
    Now, $D_{1-i}$ can only be produced if there is at least one $(1-i)$-voter (distinct from $D_{1-i}$) present.
    But such a molecule does not occur in any $\vo' \in \post_\calD(\vo)$
    and therefore also does not occur in any $\vec{d}' \in \post_{\calD'}(\vec{d})$.
}

\opt{normal}{\begin{Proof}\proofLemNoEmptyProd\end{Proof}}

\section{Generalized chemical reaction deciders} \label{sec:gcrd}

In this section, we formulate a more generalized definition of CRDs that captures
the original consensus-based definition (con-CRD) in Section~\ref{ssec:crds} and
the new existence-based definition (exi-CRD) in Section~\ref{sec:asym_os},
as well as the ``democratic'' definition (dem-CRD) in Section~\ref{sec:dem-crd}.
In this section we show how to use a result of~\cite{DBLP:conf/concur/EsparzaGLM15,Esparza2016} to re-prove the result of Angluin, Aspnes, and Eisenstat~\cite{dblp:conf/podc/angluinae06} that con-CRDs decide only semilinear sets.
This is a warmup to our main results, shown in Sections~\ref{sec:asym_os} and~\ref{sec:dem-crd}, that exi-CRDs and dem-CRDs decide exactly the semilinear sets.

\newcommand{\figGCRDWidth}{0.7\textwidth}
\begin{figure}[ht]
    \centering
    \includegraphics[width=\figGCRDWidth]{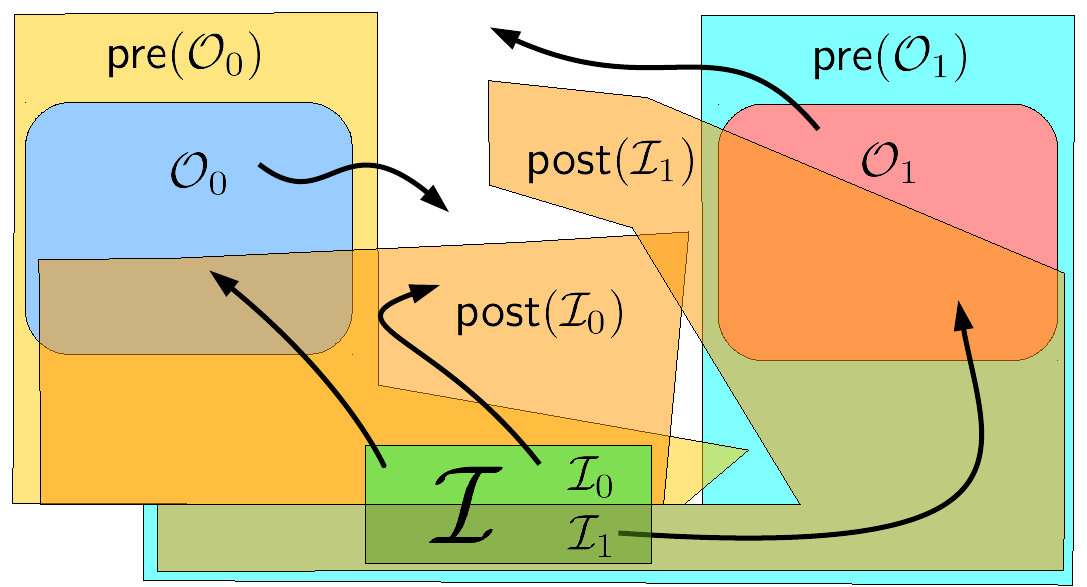}
    \caption{\figureSize Venn diagram of configurations that define generalized chemical reaction decider (gen-CRD).
    Like con-CRD, $\pre(\calO_0)$ and $\pre(\calO_1)$ partition the input set $\calI = \calI_0 \cup \calI_1$.
    The arrows, again depicting possible trajectories in the set of configurations when reactions take place, illustrate important differences with con-CRD:
    (1) possibly $\calO_i \subsetneq \post(\calO_i)$ (output is not necessarily ``stable'') and
    (2) although $\calI_i \subseteq \pre(\calO_i)$ (correct output reachable initially),
    yet possibly $\post(\calI_i) \not \subseteq \pre(\calO_i)$ (correct output could become unreachable).
    }
    \label{fig:gcrd-defn}
\end{figure}

In the generalized notion defined below we have dropped the specific structure of $\calI$, $\calO_0$, and $\calO_1$ (they are now arbitrary subsets of $\N^\Lambda$) and we have replaced the requirement that $\post(\calI_i) \subseteq \pre(\calO_i)$ by the weaker condition that $\calI_i = \calI \cap \pre(\calO_i)$ (recall Remark~\ref{rem:weaker-calI}).
Also, we do not use the term ``stable'' in reference to this generalized notion, since there is no requirement that the sets of output configurations $\calO_i$ are closed under application of reactions (i.e., we allow $\calO_i \subsetneq \post(\calO_i)$).

The relationships among the sets relevant to the definition below are illustrated in Figure~\ref{fig:gcrd-defn}.

\begin{definition} \label{def:gcrd}
A \emph{generalized chemical reaction decider (gen-CRD)}  is a $4$-tuple $\calD = (\calN,\calI,\calO_0,\calO_1)$, where $\calN=(\Lambda,R)$ is a CRN,
$\calI,\calO_0,\calO_1 \subseteq \N^\Lambda$, and there is a partition $\{\calI_0, \calI_1\}$ of $\calI$ such that $\calI_i = \calI \cap \pre(\calO_i)$ for $i \in \{0,1\}$.
\end{definition}

Observe that every con-CRD is a gen-CRD.
However, the requirements to be a gen-CRD are weaker than for con-CRDs:
(1) the condition $\post(\calO_i) = \calO_i$ need not hold for gen-CRDs, so it may be possible to ``escape'' from $\calO_i$, and
(2) since $\post(\calI_i) \subseteq \pre(\calO_i)$ need not hold for gen-CRDs, it is possible to take a ``wrong'' route starting from $\calI_i$ such that $\calO_i$ becomes unreachable.\footnote{While Definition~\ref{def:gcrd} appears almost too general to be useful,
Corollary~\ref{cor:semil_GRCD_decide_semil} says that if $\calI, \calO_0, \calO_1$ are semilinear, then so are $\calI_0,\calI_1$,
which implies that any CRD definition that can be framed as such a gen-CRD must decide only semilinear sets.}

\begin{example}
Consider again the con-CRD $\calD = (\calN,\calI,\calO_0,\calO_1)$ from Example~\ref{ex:symCRD} (for some fixed constant $m \geq 2$). Let $\calN' = (\Lambda \cup \{G\},R \cup \{\alpha\})$ be the CRN obtained from $\calN = (\Lambda,R)$ by adding a new species $G \notin \Lambda$ and adding the reaction $\alpha = X \to X+G$. Also, let $\calI',\calO'_0,\calO'_1$ be obtained from $\calI,\calO_0,\calO_1$, respectively, by padding for each configuration a zero entry for species $G$. Then $\calD' = (\calN',\calI',\calO'_0,\calO'_1)$ is a gen-CRD where $\post(\calI_i) \not\subseteq \pre(\calO_i)$ --- indeed, once reaction $\alpha$ has taken place we cannot reach any $\calO_i$. So, we have taken a ``wrong'' route once reaction $\alpha$ has taken place at least once. We also have $\post(\calO_1) \neq \calO_1$ since there are configurations of $\calO_1$ for which reaction $\alpha$ can take place and once $\alpha$ has taken place we are outside $\calO_1$.
\end{example}

Despite these relaxations,
observe that the following property of con-CRDs is retained in gen-CRDs:
$\calI$ is the disjoint union of $\calI_0 = \calI \cap \pre(\calO_0)$ and $\calI_1 = \calI \cap \pre(\calO_1)$, \emph{i.e.}, from each input configuration, \emph{exactly one} of the two output sets $\calO_0$ or $\calO_1$ is reachable.
We say that a gen-CRD $\calD$ \emph{decides} the set $\calI_1$.

Definition~\ref{def:gcrd} is inspired by the following key Petri net result from \cite[Theorem~10]{Esparza2016} (announced in \cite[Theorem~10]{DBLP:conf/concur/EsparzaGLM15}), formulated here in terms of CRNs.

\begin{theorem}[\cite{DBLP:conf/concur/EsparzaGLM15,Esparza2016}] \label{thm:concur_petri_result}
Let $\calN$ be a CRN and $\calI,\calO_0,\calO_1 \subseteq \N^\Lambda$ be semilinear. Let $\calI_i = \calI \cap \pre(\calO_i)$ for $i \in \{0,1\}$.
If $\{\calI_0, \calI_1\}$ is a partition of $\calI$, then $\calI_0$ and $\calI_1$ are semilinear.
\end{theorem}

We say that a gen-CRD $\calD = (\calN,\calI,\calO_0,\calO_1)$ is \emph{semilinear} if $\calI$, $\calO_0$, and $\calO_1$ are all semilinear. We immediately have the following corollary to Theorem~\ref{thm:concur_petri_result}.

\begin{corollary} \label{cor:semil_GRCD_decide_semil}
If a semilinear gen-CRD decides $X \subseteq \N^\Lambda$, then $X$ is semilinear.
\end{corollary}

As a by-product of the results shown in \cite{DBLP:conf/concur/EsparzaGLM15,Esparza2016},
the reverse direction of Theorem~\ref{thm:ostable_semilinear} (which is the most difficult implication) was reproven in \cite{DBLP:conf/concur/EsparzaGLM15,Esparza2016} for the case of population protocols.
That proof however essentially uses the fact that, for population protocols, $\post(\vc)$ is finite for all configurations $\vc$, which is not true for CRNs in general.
Fortunately, one may still obtain the full reverse direction of Theorem~\ref{thm:ostable_semilinear} by showing that every con-CRD is semilinear (cf.\ the proof of Theorem~\ref{thm:ostable_implies_semilinear} below) and then invoking Corollary~\ref{cor:semil_GRCD_decide_semil}.

We now use this machinery to re-prove the result, due originally to Angluin, Aspnes, and Eisenstat~\cite{dblp:conf/podc/angluinae06}, that con-CRDs decide only semilinear sets.

\begin{theorem} \label{thm:ostable_implies_semilinear}
Every con-CRD decides a semilinear set.
\end{theorem}

\begin{Proof}
Let $\calD = (\calN,\calI,\calO_0,\calO_1)$ be a con-CRD.
Let $\calI' = \{ \vc \in \N^\Lambda \mid \vc\rest{\Lambda \setminus \Sigma} = \vec{0} \}$.
The complement of $\calI'$ is closed upwards, thus $\calI'$ is semilinear, as is $\calI = \calI' \setminus \{ \vec{0}\}$.

We now show that each $\calO_i$ is semilinear.
Let $\calL_i = \{ \vc \in \N^\Lambda \mid \vc\rest{\Gamma_i} \neq \vec{0} \}$ as in Definition~\ref{def:os_crd}.
By Remark~\ref{rem:Oi-pre-Lotheri}, $\calO_i = \N^\Lambda \setminus \pre(\calL_{1-i} \cup \{\vec{0}\}) = \N^\Lambda \setminus (\pre(\calL_{1-i}) \cup \pre(\vec{0}))$.
By Lemma~\ref{lem:no_empty_prod} we may assume that each reaction $\alpha = (\vr,\vp)$ of $\calD$ has $\vp \neq \vec{0}$, so $\pre(\vec{0})=\{\vec{0}\}$, which is semilinear.
Since $\calL_{1-i}$ is closed upwards, by Lemma~\ref{lem:closed_upw_pre_post}, $\pre(\calL_{1-i})$ is also closed upwards, so semilinear by Lemma~\ref{lem:closed_upw_semil}.
Since semilinear sets are closed under union and complement, $\calO_i$ is also semilinear, so $\calD$ is a semilinear gen-CRD.
The theorem follows by Corollary~\ref{cor:semil_GRCD_decide_semil}.
\end{Proof}

\begin{remark} \label{rem:post-sub-pre-unnecessary}
From the hypothesis $\post(\calI_i) \subseteq \pre(\calO_i)$ in Definition~\ref{def:os_crd},
we used only the weaker conclusion $\calI_i = \calI \cap \pre(\calO_i)$.
In other words, we need merely that $\calO_i$ is \emph{initially} reachable from $\calI_i$ itself (\emph{and} that $\calO_{1-i}$ is unreachable from $\calI_i$, since $\pre(\calO_0)$ and $\pre(\calO_{1})$ partition $\calI$).
We do not require that $\calO_i$ \emph{remains} reachable from every configuration reachable from $\calI_i$ (i.e., $\post(\calI_i)$).
Hence one could weaken part~\ref{defn:o-stable-crd-preI-in-postO} of Definition~\ref{def:os_crd} to use the condition
$\calI_i = \calI \cap \pre(\calO_i)$,
and Theorem~\ref{thm:ostable_implies_semilinear} still holds.\footnote{In contrast, the proof of~\cite{dblp:conf/podc/angluinae06} crucially requires the hypothesis
$\post(\calI_i) \subseteq \pre(\calO_i)$.}
\end{remark}

Despite Remark~\ref{rem:post-sub-pre-unnecessary}, if a gen-CRD \emph{does} obey the stronger condition $\post(\calI_i) \subseteq \pre(\calO_i)$, then a convenient property holds:
each $\calO_i$ may be enlarged without altering the set $\calI_1$ decided by the gen-CRD, so long as $\calO_{1-i}$ remains unreachable from $\calO_i$.
The following lemma formalizes this.

\begin{lemma} \label{lem:os_blowup}
Let $\calD = (\calN,\calI,\calO_0,\calO_1)$ be a gen-CRD that decides $\calI_1$ and let $\calI_0 = \calI \setminus \calI_1$.
For $i \in \{0,1\}$, assume that $\post(\calI_i) \subseteq \pre(\calO_i)$,
and let $\calO'_i \supseteq \calO_i$ with
$\post(\calO'_i) \cap \calO_{1-i} = \emptyset$.
Then $\calD' = (\calN,\calI,\calO'_0,\calO'_1)$ is a gen-CRD deciding $\calI_1$.
\end{lemma}

\begin{Proof}
We have $\calI_i = \pre(\calO_i) \cap \calI \subseteq \pre(\calO'_i) \cap \calI$ for $i \in \{0,1\}$.
To show that this inclusion is an equality, it suffices to show that $\pre(\calO'_0) \cap \calI$ and $\pre(\calO'_1) \cap \calI$ are disjoint.

Let $\vi \in \calI_i$. Then $\vi \in \pre(\calO_i) \subseteq \pre(\calO'_i)$.
Assume to the contrary $\vi \in \pre(\calO'_{1-i})$.
Let $\vo \in \calO'_{1-i} \cap \post(\vi)$,
so  $\vo \in \post(\vi) \subseteq \post(\calI_i) \subseteq \pre(\calO_i)$.
Thus $\calO'_{1-i} \cap \pre(\calO_i) \neq \emptyset$.
In other words, $\post(\calO'_{1-i}) \cap \calO_i \neq \emptyset$ --- a contradiction.
Hence $\pre(\calO'_0) \cap \calI$ and $\pre(\calO'_1) \cap \calI$ are disjoint.
\end{Proof}

\section{Existential output-stability} \label{sec:asym_os}

We now give a natural alternative output convention for CRDs, which we call an existential output-stable CRD (exi-CRD).
Whereas the output $i$ of a con-CRD is based on both the presence of species of one type $\Gamma_i$ and the absence of a species of a different type $\Gamma_{1-i}$, the output of an exi-CRD is based solely on the presence or absence of a single species type $\Gamma_1$.

For each $\vi \in \calI$ the CRD can either
(1) reach a configuration $\vo$ so that for each configuration $\vo'$ reachable from $\vo$ (including $\vo$ itself) we have $\vo'\rest{\Gamma_1} \neq \vec{0}$ or
(2) reach a configuration $\vo$ so that for each configuration $\vo'$ reachable from $\vo$ we have $\vo'\rest{\Gamma_1} = \vec{0}$.
Similarly to gen-CRDs, and unlike con-CRDs,\footnote{As noted, con-CRDs could be defined by replacing the requirement $\post(\calI_i) \subseteq \pre(\calO_i)$ with $\calI_i = \calI \cap \pre(\calO_i)$ and retain the same power, but for clarity we retain the original definition.}
it is not required that such a configuration $\vo$ is reachable from \emph{any} configuration $\vc$ reachable from the initial $\vi$, merely that such a $\vo$ is reachable from $\vi$ itself.
Even this more liberal assumption does not allow the CRD to decide a non-semilinear set.

\begin{definition}\label{def:aos_crd}
An \emph{existential output-stable chemical reaction decider (exi-CRD)} is a gen-CRD
$\calD = (\calN,\calI,\calO_0,\calO_1)$,
where there are $\Sigma \subseteq \Lambda$ and voting species $\Gamma_1 \subseteq \Lambda$ such that
\begin{enumerate}
    \item \label{defn:asym-CRD-cond1}
    $\calI = \{ \vc \in \N^\Lambda \mid \vc\rest{\Lambda\setminus\Sigma} = \vec{0} \} \setminus \{\vec{0}\}$, and

    \item \label{defn:asym-CRD-cond2}
    $\calO_i = \{ \vc \in \N^\Lambda \mid \post(\vc) \subseteq \mathcal{V}_i \}$ for $i \in \{0,1\}$, with $\mathcal{V}_1 = \{ \vc \in \N^\Lambda \mid \vc\rest{\Gamma_1} \neq \vec{0} \}$ and $\mathcal{V}_0 = \N^\Lambda \setminus \mathcal{V}_1$.\footnote{Just as for con-CRDs,  $\post(\calO_i) = \calO_i$.
    Note that $\calV_1$ above is the same as $\calL_1$ in Definition~\ref{def:os_crd}, but $\calL_0 \neq \calV_0$, since $\calL_1$ and $\calL_0$ can have nonempty intersection if there are conflicting voters present in some configuration.}
\end{enumerate}
\end{definition}

Condition~\ref{defn:asym-CRD-cond1} states that only species in $\Sigma$ may be present initially, and at least one must be present.
Condition~\ref{defn:asym-CRD-cond2} defines $\mathcal{V}_1$ and $\mathcal{V}_0$ to be configurations with and without $\Gamma_1$ voters, and $\calO_i$ to be the stable subsets of $\mathcal{V}_i$.

\begin{example}
Consider the following exi-CRD $\calD'$, where $\Lambda = \Sigma = \Gamma_1 = \{X,Y\}$,
which decides the same set as in Example~\ref{ex:symCRD} (i.e., $x \not\equiv y \mod m$).
\begin{eqnarray}
mX \to \emptyset, \quad mY \to \emptyset, \quad X+Y \to \emptyset.  \label{eq:asymCRD}
\end{eqnarray}
If $x \equiv y \mod m$, then eventually all $X$ and $Y$ molecules are consumed and we obtain the configuration $\vc  = \vec{0} \in \calO_0$.
Otherwise, all $X$ and $Y$ molecules cannot be consumed, and we are in $\calO_1$.
This example illustrates that the exi-CRD computing convention may permit a simpler implementation in some cases.
Indeed, compared with Example~\ref{ex:symCRD}, (\ref{eq:asymCRD}) has $2$ fewer reactions and $1$ fewer species (and is also faster since fewer reactions need to occur).
\end{example}

We first observe that exi-CRDs have at least the computational power of con-CRDs.

\begin{observation} \label{obs:ostable_leq_aostable}
Let $\calD=(\calN,\calI,\calO_0,\calO_1)$ be a con-CRD deciding $X$, with voter partition $\{\Gamma_0,\Gamma_1\}$.
Then $\calD'=(\calN,\calI,\calO_0',\calO_1')$, where, for $i \in \{0,1\}$, $\calO_i' = \{ \vc \in \N^\Lambda \mid \post(\vc) \subseteq \mathcal{V}_i \}$, with $\calV_i$ as in Definition~\ref{def:aos_crd} (with respect to $\Gamma_1$), is an exi-CRD deciding $X$.
\end{observation}

\begin{Proof}
This follows from Lemma~\ref{lem:os_blowup} since (1) $\calO_i \subseteq \calO'_i$ and (2) $\post(\calO'_i) = \calO'_i$ is disjoint from $\calO_{1-i}$ for $i \in \{0,1\}$.
\end{Proof}

We now show that exi-CRDs have \emph{no greater} computational power than con-CRDs.
This is not as immediate as the other direction.
First, observe that an exi-CRD may not be a con-CRD; if we interpret species $V_0 \in \Lambda \setminus \Gamma_1$ as voting ``0'', then a con-CRD is required to \emph{eliminate} them to output ``1'', but not an exi-CRD.
Moreover, a direct transformation of an exi-CRD into a con-CRD appears difficult.
Intuitively, the problem is that the absence of molecules in $\Gamma_1$ is not detectable by a CRN, so there is no obvious way to ensure that a species $V_0 \in \Lambda \setminus \Gamma_1$ is produced only if all $V_1 \in \Gamma_1$ are absent.
The next obvious proof strategy would be to show, as in the proof of Theorem~\ref{thm:ostable_implies_semilinear},
that every exi-CRD is a semilinear gen-CRD.
However, it is not clear whether $\calO_1$ is semilinear.
Nonetheless, due to the generality of Definition~\ref{def:gcrd} and Theorem~\ref{thm:concur_petri_result}, we can define a semilinear gen-CRD that decides the same set, by taking a subset of $\calO_1$ that is provably semilinear and still satisfies the necessary reachability constraints, even though the gen-CRD we define is \emph{not} in fact an exi-CRD (in particular, its ``output'' set $\calO_1$ is not closed under application of reactions).

\newcommand{\nondec}{\mathsf{nondec}}

Recall that a \emph{homomorphism} $f:\N^\Lambda\to\Z$ obeys $f(\vc+\vc') = f(\vc)+f(\vc')$ for all $\vc,\vc' \in \N^\Lambda$.
Some examples include $f(\vc) = \vc(S)$ for some $S\in\Lambda$,
$f(\vc) = \norm{\vc\rest\Delta}$ for some $\Delta \subseteq \Lambda$,
or $f(\vc) = \vc(S_1)-\vc(S_2)$ for some $S_1,S_2\in\Lambda$.

For a CRN $\calN$ and a function $f: \N^\Lambda \to \Z$,
we define $\nondec_{f,\calN} = \{ \vc \in \N^\Lambda \mid \forall \vc' \in \post(\vc), f(\vc') \geq f(\vc) \}$ as the set of configurations $\vc$
in which $f$ is minimal among all the configurations reachable from $\vc$.

We now prove a key lemma, which will be used for characterizing both exi-CRDs in this section and dem-CRDs in Section~\ref{sec:dem-crd}.

\begin{lemma} \label{lem:nondec}
Let $\calN$ be a CRN and $f: \N^\Lambda \to \Z$ a homomorphism.
Let  $\calO = \{ \vc \in \N^\Lambda \mid \post(\vc) \subseteq \mathcal{V} \}$
with $\mathcal{V} = \{ \vc \in \N^\Lambda \mid f(\vc) > 0 \}$.
Then $\calO \cap W$ is semilinear and $\pre(\calO \cap W) = \pre(\calO)$, where $W = \nondec_{f,\calN}$.
\end{lemma}

\begin{Proof}
We first prove $\pre(\calO \cap W) = \pre(\calO)$.
Obviously, $\pre(\calO \cap W) \subseteq \pre(\calO)$.
To prove the reverse containment, let $\vc \in \pre(\calO)$.
Hence $\vc \in \pre(\vo)$ for some $\vo \in \calO$.
Since every $\vo' \in \post(\vo)$ satisfies $f(\vo')>0$,
there is an $\vo' \in \post(\vo)$ such that $f(\vo')$ is minimal among all configurations in $\post(\vo)$.
Thus $\vo' \in W$.
Since $\post(\calO) = \calO$, we have $\vo' \in \calO$.
Hence, $\vo' \in \calO \cap W$.
Now, $\vo \in \pre(\vo')$ and $\vc \in \pre(\vo)$, and so $\vc \in \pre(\vo')$.
Therefore, $\vc \in \pre(\calO \cap W)$, so $\pre(\calO) \subseteq \pre(\calO \cap W)$.

We now show that $\calO \cap W$ is semilinear.
Observe that the set $\N^\Lambda \setminus W = \{ \vc \in \N^\Lambda \mid \exists \vc' \in \post(\vc), f(\vc') < f(\vc) \}$ is closed upwards. Indeed, if $\vc \in \N^\Lambda \setminus W$ and $\vc' \in \post(\vc)$ with $f(\vc') < f(\vc)$, then for all $\vec{d} \in \N^\Lambda$, $\vc'+\vec{d} \in \post(\vc+\vec{d})$ and $f(\vc'+\vec{d}) = f(\vc') + f(\vec{d}) < f(\vc) + f(\vec{d}) = f(\vc + \vec{d})$.
Thus $\N^\Lambda \setminus W$ is semilinear by Lemma~\ref{lem:closed_upw_semil}, and
hence also $W$.
Since $\calO \subseteq \mathcal{V}$, we have $\calO \cap W \subseteq \mathcal{V} \cap W$.
Conversely, if $\vc \in \mathcal{V} \cap W$,
then $f(\vc) > 0$ since $\vc \in \calV$,
and for all $\vc' \in \post(\vc)$, $f(\vc') \geq f(\vc) > 0$ since $\vc \in W$.
Thus $\vc \in \calO \cap W$, showing $\calO \cap W = \mathcal{V} \cap W$, which is semilinear since $\mathcal{V}$ and $W$ are.
\end{Proof}

Using Lemma~\ref{lem:nondec} we show that every exi-CRD can be changed into a semilinear gen-CRD by choosing $\calO_1 \cap W$, rather than $\calO_1$, as its ``output 1'' set of configurations.
Note that unlike in the definition of con-CRD and exi-CRD, $\calO_1 \cap W$ is \emph{not} in general closed under application of reactions.

\begin{lemma} \label{lem:alostable_implies_semilinear}
Let $\calD = (\calN,\calI,\calO_0,\calO_1)$ be an exi-CRD deciding $X$ and $\Gamma_1$ be as in Definition~\ref{def:aos_crd}.
Let $W = \nondec_{f,\calN}$ with $f: \N^\Lambda \to \Z$ defined as $f(\vc) = \norm{\vc\rest{\Gamma_1}}$ for all $\vc \in \N^\Lambda$.
Then $\calD' = (\calN,\calI,\calO_0,\calO_1 \cap W)$ is a semilinear gen-CRD deciding $X$.
\end{lemma}

\begin{Proof}
Observe that $f$ is a homomorphism. Now, Lemma~\ref{lem:nondec} tells us that $\pre(\calO_1 \cap W) = \pre(\calO_1)$; thus $\calD'$ decides $X$.

To complete the proof, it suffices to show that $\calD'$ is semilinear.
$\calI$ is obtained from the closed-upwards set $\N^\Sigma \setminus \{\vec{0}\}$ by padding zeros for the species of $\Lambda \setminus \Sigma$, so $\calI$ is semilinear. $\calO_1 \cap W$ is semilinear by Lemma~\ref{lem:nondec}.
To see that $\calO_0$ is semilinear,
let $\mathcal{V}_0$ and $\mathcal{V}_1$ be as in Definition~\ref{def:aos_crd}.
Clearly $\mathcal{V}_1$ is closed upwards, so semilinear.
So, (1) $\pre(\mathcal{V}_1)$ is also closed upwards and therefore semilinear (by Lemma~\ref{lem:closed_upw_pre_post} and Lemma~\ref{lem:closed_upw_semil}) and
(2) $\mathcal{V}_0 = \N^\Lambda \setminus \mathcal{V}_1$ is semilinear.
Thus, $\calO_0 = \mathcal{V}_0 \setminus \pre(\mathcal{V}_1)$ is semilinear since the class of semilinear sets is closed under set difference.
\end{Proof}

The following is the first of two main results of this paper.
It says that the computational power of con-CRDs equals that of exi-CRDs; they both decide exactly the semilinear sets.

\begin{theorem}\label{thm:aostable_semilinear}
Let $X \subseteq \N^\Sigma \setminus \{\vec{0}\}$. Then $X$ is semilinear if and only if there is an exi-CRD that decides $X$.
\end{theorem}

\begin{Proof}
The forward direction follows from Observation~\ref{obs:ostable_leq_aostable} and Theorem~\ref{thm:ostable_semilinear}.
For the reverse direction, let $\calD$ be an exi-CRD deciding $X$.
By Lemma~\ref{lem:alostable_implies_semilinear}, there is a semilinear gen-CRD $\calD'$ deciding $X$,
which is semilinear by Corollary~\ref{cor:semil_GRCD_decide_semil}.
\end{Proof}

\section{Democratic output-stability}\label{sec:dem-crd}

Another reasonable alternative output convention is the one most naturally associated with the term ``voting'': a \emph{democratic} output convention in which, rather than requiring a consensus, we define output by majority vote.
In this case, for sets of voting species $\Gamma_0$ and $\Gamma_1$, the only undefined outputs occur in ``tie'' configurations $\vc$ where $\norm{\vc\rest{\Gamma_0}} = \norm{\vc\rest{\Gamma_1}}$.
In this section we show that such CRDs have equivalent computing power to con-CRDs.

\begin{definition}\label{defn:dem-crd}
  A \emph{democratic output-stable chemical reaction decider (dem-CRD)} is a gen-CRD
  $\calD=(\calN,\calI,\calO_0,\calO_1)$,
  where there are $\Sigma \subseteq \Lambda$ and
  a partition $\{\Gamma_0,\Gamma_1\}$ of $\Lambda$
  such that
\begin{enumerate}
    \item \label{defn:dem-crd-I}
    $\calI = \{ \vc \in \N^\Lambda \mid \vc\rest{\Lambda\setminus\Sigma} = \vec{0} \} \setminus \{\vec{0}\}$,

    \item \label{defn:dem-crd-O}
    $\calO_i = \{ \vc \in \N^\Lambda \mid \post(\vc) \subseteq \calM_i \}$,
    with
    $\calM_i = \{ \vc \in \N^\Lambda \mid \norm{\vc\rest{\Gamma_i}} > \norm{\vc\rest{\Gamma_{1-i}}} \}$ for $i \in \{0,1\}$.

\end{enumerate}
\end{definition}

Note that $\calM_0 \cap \calM_1 = \emptyset$, and
that $\calO_i$ is stable, i.e., $\calO_i = \post(\calO_i)$.
A con-CRD reaches a consensus, the strongest kind of majority, leading to the following observation implying that dem-CRDs are at least as powerful as con-CRDs.

\begin{observation}\label{obs:sym-crd-to-dem-crd}
Let $\calD=(\calN,\calI,\calO_0,\calO_1)$ be a con-CRD deciding $X$, with voter partition $\{\Gamma_0,\Gamma_1\}$.
Then $\calD'=(\calN,\calI,\calO_0',\calO_1')$, where $\calO_i' = \{ \vc \in \N^\Lambda \mid \post(\vc) \subseteq \mathcal{M}_i \}$ for $i \in \{0,1\}$, with $\calM_i$ as in Definition~\ref{defn:dem-crd}, is a dem-CRD deciding $X$.
\end{observation}

\begin{Proof}
This follows from Lemma~\ref{lem:os_blowup} since (1) $\calO_i \subseteq \calO'_i$ and (2) $\post(\calO'_i) = \calO'_i$ is disjoint from $\calO_{1-i}$ for $i \in \{0,1\}$.
\end{Proof}

\newcommand{\thmDemCRDSemilinear}{
    Let $X \subseteq \N^\Sigma \setminus \{\vec{0}\}$.
    Then $X$ is semilinear if and only if there is a dem-CRD that decides $X$.
}

The converse result, that dem-CRDs are no more powerful than con-CRDs, implies the second main result of this paper.
\opt{sub}{
The proof of the following theorem is found in the full version of this paper, and relies on the gen-CRD framework of Section~\ref{sec:gcrd} and Lemma~\ref{lem:nondec} (choosing $f$ that is the difference between $0$ and $1$ voters).
}

\begin{theorem}\label{thm:dem-crd-semilinear}
\thmDemCRDSemilinear
\end{theorem}

\newcommand{\argumentDemCRDSemilinear}{
    In order to prove Theorem~\ref{thm:dem-crd-semilinear}, we first show the following lemma.

    \begin{lemma} \label{lem:dem_implies_semilinear}
    Let $\calD = (\calN,\calI,\calO_0,\calO_1)$ be a dem-CRD that decides $X$ and $\calM_i$ for $i\in\{0,1\}$ be as in Definition~\ref{defn:dem-crd}.
    Let, for $i \in \{0,1\}$, $W_i = \nondec_{f_i,\calN}$ with $f_i: \N^\Lambda \to \Z$ such that $f_i(\vc) = \norm{\vc\rest{\Gamma_i}}-\norm{\vc\rest{\Gamma_{1-i}}}$ for all $\vc \in \N^\Lambda$.
    Then $\calD' = (\calN,\calI,\calO_0 \cap W_0,\calO_1 \cap W_1)$ is a semilinear gen-CRD deciding $X$.
    \end{lemma}

    \begin{Proof}
    Let $i\in\{0,1\}$. Observe that $f_i$ is a homomorphism.
    Lemma~\ref{lem:nondec} says that $\pre(\calO_i \cap W_i) = \pre(\calO_i)$, so $\calD'$ decides $X$.
    To see that $\calD'$ is semilinear,
    note that $\calI$ is semilinear, and for $i\in\{0,1\}$,
    $\calO_i \cap W_i$ is semilinear by Lemma~\ref{lem:nondec}.
    \end{Proof}

    We are now ready to prove Theorem~\ref{thm:dem-crd-semilinear}.

    \opt{sub}{\rethm{\ref{thm:dem-crd-semilinear}}{\thmDemCRDSemilinear}}

    \opt{normal}{\begin{Proof}[of Theorem~\ref{thm:dem-crd-semilinear}]}
    \opt{sub}{\begin{Proof}}
    The forward direction follows from Observation~\ref{obs:sym-crd-to-dem-crd} and Theorem~\ref{thm:ostable_semilinear}.
    For the reverse direction, let $\calD$ be a dem-CRD deciding $X$.
    By Lemma~\ref{lem:dem_implies_semilinear}, there is a semilinear gen-CRD $\calD'$ deciding $X$,
    which is semilinear by Corollary~\ref{cor:semil_GRCD_decide_semil}.
    \end{Proof}
}

\opt{normal}{\argumentDemCRDSemilinear}

\section{Discussion} \label{sec:discussion}
Using a recent result about Petri nets~\cite{DBLP:conf/concur/EsparzaGLM15,Esparza2016} (cf.\ Theorem~\ref{thm:concur_petri_result}) we have presented a framework able to capture different output conventions for computational CRNs.
The original consensus-based definition~\cite{dblp:journals/dc/angluinadfp06} can be fitted in this framework,
giving a new proof that such CRNs are limited to computing only semilinear sets.
Two additional definitions, an existence-based convention, and a  majority-vote convention, can be fitted in this framework, and thus have the same expressive power as the original.

We show that exi-CRDs and dem-CRDs are no more powerful than con-CRDs by showing that they are limited to deciding semilinear sets, which is known also to apply to con-CRDs.
It would be informative, however, to find a proof that uses a direct simulation argument, showing how to transform an arbitrary exi-CRD or dem-CRD into a con-CRD deciding the same set.
Along a similar line of thinking, we have defined the computational ability of CRDs without regard to time complexity, which is potentially sensitive to definitional choices, even if the class of decidable sets remains the same~\cite{angluin2008fast,DBLP:conf/dna/DotyH13,LeaderElectionDISC,TimeSpaceTradeoffsPP,polylogleaderICALP}.
It would be interesting to find cases in which exi-CRDs or dem-CRDs are be able to compute faster than any equivalent con-CRD.

An open problem is to consider other output conventions, where we possibly step out of semilinearity.
For example, consider a designated species $V_1$ such that for each input configuration $\vec{d} \in \calI$,
(1) $\vec{d} \in \calI_1$ if we always eventually reach a configuration $\vc$ such that all configurations reachable from $\vc$ has a $V_1$ molecule, and
(2) $\vec{d} \in \calI_0$ if we can never reach such a configuration $\vc$.
Hence the output of a configuration is then based on a behavioral property of the system (whether it is stable)
instead of a syntactic property of the configuration (whether it contains a particular molecule).
It is not clear how to apply Theorem~\ref{thm:concur_petri_result}, which requires that $\calI_0 = \calI \cap \pre(S)$ for some semilinear set $S$.

\opt{normal}{
It would be interesting to find generalizations of Theorem~\ref{thm:concur_petri_result} beyond semilinearity of the sets $\calI,\calO_0,\calO_1$, showing that if they satisfy some condition, then so do $\calI_0$ and $\calI_1$.

In addition to predicates (functions with \emph{binary} output), computation by CRNs computing \emph{integer}-valued functions has also been extensively investigated~\cite{CheDotSolNaCo, DBLP:conf/dna/DotyH13, DBLP:conf/innovations/2014, ProgrCRNs/winfree_solo, DBLP:journals/nc/SoloveichikCWB08, DBLP:conf/dna/CummingsDS14}.
It remains to investigate alternative output conventions for such functions, and in particular how composable such conventions are with each other, since the output of a function $f: \N \to \N$ can be the input of another function $g:\N \to \N$.
}

\newcommand{\acks}{
\paragraph{Acknowledgements.}
R.B.\ thanks Grzegorz Rozenberg for useful comments on an earlier version of this paper and for useful discussions regarding CRNs in general.
D.D.\ thanks Ryan James for suggesting the democratic CRD model.
The authors are grateful to anonymous reviewers for comments on a conference version of this paper and on an earlier version of this journal paper that have helped improve the presentation.
}
\acks

\bibliographystyle{plain}
\bibliography{crns_petri}

\newpage

\appendix

\section{Consensus-based CRDs with nonvoters}\label{sec:sym-nonvoters}

A slightly modified definition of a con-CRD is found in the literature~\cite{CheDotSolNaCo},
in which only a subset of species is designated as voters, and nonvoting species do not affect the output.
Unlike exi-CRDs, which also have only a subset of voting species, these CRDs treat ``yes'' and ``no'' votes symmetrically with respect to interpreting what is the ``output'' of a configuration.
We refer to this as a \emph{delegating} CRD (in analogy to \emph{delegates} who vote on behalf of others).

\begin{definition}\label{def:del_crd}
  A \emph{delegating output-stable chemical reaction decider (del-CRD)} is a gen-CRD $\calD=(\calN,\calI,\calO_0,\calO_1)$
  where $\calN=(\Lambda,R)$ is a CRN and there are $\Sigma \subseteq \Lambda$ and disjoint subsets of voting species $\Gamma_0,\Gamma_1 \subseteq \Lambda$ such that
\begin{enumerate}
    \item \label{defn:del-sym-crd-I}
    $\calI = \{ \vc \in \N^\Lambda \mid \vc\rest{\Lambda\setminus\Sigma} = \vec{0} \} \setminus \{\vec{0}\}$,

    \item \label{defn:del-sym-crd-O}
    $\calO_i = \{ \vc \in \N^\Lambda \mid \post(\vc) \subseteq \calL_i \setminus \calL_{1-i} \}$,
    with
    $\calL_i = \{ \vc \in \N^\Lambda \mid \vc\rest{\Gamma_i} \neq \vec{0} \}$ for $i \in \{0,1\}$.

    \item \label{defn:del-sym-crd-preI-in-postO}
    There is a partition $\{\calI_0,\calI_1\}$ of $\calI$ such that $\post(\calI_i) \subseteq \pre(\calO_i)$ for $i \in \{0,1\}$.

\end{enumerate}
\end{definition}

The only difference between a con-CRD and a del-CRD is that the latter omits the requirement that $\Gamma_0 \cup \Gamma_1 = \Lambda$, so each con-CRD is a del-CRD.
To show they have equivalent computational power, it then suffices to show that any del-CRD can be turned into a con-CRD deciding the same set.
This equivalence is simpler to establish than for exi-CRDs and dem-CRDs, using a direct simulation argument that does not require the machinery of gen-CRDs.

\begin{lemma} \label{lem:del-sym-crd-to-sym-crd}
  For each del-CRD, there is a con-CRD deciding the same set.
\end{lemma}

\begin{Proof}
Let $\calD =  (\calN,\calI,\calO_0,\calO_1)$ be an del-CRD deciding $X$, with $\calN=(\Lambda,R)$ and voting species $\Gamma_0,\Gamma_1 \subseteq \Lambda$ as in Definition~\ref{def:del_crd}.
Let $\Delta = \Lambda \setminus (\Gamma_0 \cup \Gamma_1)$ be the nonvoting species.
Intuitively, we define a CRN $\calN'$ in which all nonvoting species $S \in \Delta$ of $\calN$ have an additional bit that determines whether $S$ is a $0$-voter or a $1$-voter.
We add reactions so that species in $\Gamma_i$ flip this bit to $i$ in any molecule in $\Delta$.
More precisely, let $\calN'$ be obtained from $\calN$ by first replacing every species $S \in \Delta$ by two species $S_0$ and $S_1$.
Let $\Lambda'$ be the obtained set of species of $\calN'$.
Replace every reaction $\alpha = (\vr,\vp)$ of $\calN$
by reactions $\alpha' = (\vr',\vp')$ with $\vr',\vp' \in \N^{\Lambda'}$
such that $\pi(\vr')=\vr$ and $\pi(\vp')=\vp$,
where $\pi: \Lambda' \to \Lambda$ sends every species $S_i$ to $S$
and sends each $V_i \in \Gamma_i$ to itself (and $\pi$ is applied component-wise to vectors).
Moreover, for $i \in \{0,1\}$, add reactions $V_i+S_{1-i} \to V_i + S_i$ for all $S \in \Delta$ and $V_i \in \Gamma_i$.

Let $\calD' = (\calN',\calI',\calO'_0,\calO'_1)$, with $\calI'$, $\calO'_0$, and $\calO'_1$ defined as in Definition~\ref{def:os_crd} and $\calI'$ defined with respect to $\Sigma' = \{ S_1 \mid S \in \Sigma \}$ where $\Sigma$ corresponds to $\calI$.
(The choice of $1$ instead of $0$ is arbitrary.)
We observe that $\calD'$ is a con-CRD.
Indeed, once a configuration $\vc \in \calO_i$ in $\calD$ is reached from an input configuration, we have that for each $\vc' \in \post(\vc)$, $\vc'$ contains at least one molecule of species $V_i$ and none of $V_{1-i}$.
A configuration $\vec{d}$ in $\calD'$ corresponding to $\vc$ will turn every molecule into a $i$-voter. In other words, we eventually reach a configuration $\vec{d}' \in \calO'_i$.
Hence $\calD'$ is a con-CRD deciding $X$.
\end{Proof}

Although the converse is trivial since, in creating a del-CRD from a con-CRD, one can choose the voting species $\Gamma_0,\Gamma_1$ to be the same, in some cases it is preferable to have a strict subset.
One case in particular, in which there are exactly two voting species, i.e., $|\Gamma_0| = |\Gamma_1| = 1$, merits mention since this is often a convenient assumption to make about a CRD.
The following lemma shows that we can make this assumption without loss of generality.

\begin{lemma} \label{lem:sym-crd-to-2-voter-del-sym-crd}
  For each con-CRD, there is a del-CRD with exactly two voting species deciding the same set.
\end{lemma}

\begin{Proof}
Let $\calD = (\calN,\calI,\calO_0,\calO_1)$ be a con-CRD that decides $X$, with voting species $\Gamma_0,\Gamma_1$ that partition $\Lambda$.
Let $\calN'$ be the CRN obtained from $\calN$ by adding two new species $V_0,V_1$ to $\calD$
and adding, for each $S \in \Gamma_i$, the reactions $S \to S + V_i$ and $S + V_{1-i} \to S$.
Let $\calD' = (\calN',\calI',\calO'_0,\calO'_1)$, with $\calI'$, $\calO'_0$, and $\calO'_1$ defined as in Definition~\ref{def:del_crd} and $\calI'$ defined with respect to the same $\Sigma$.
Indeed, once an output-stable configuration $\vc \in \calO_i$ in $\calD$ is reached from an input configuration, we have that for each $\vc' \in \post(\vc)$, every molecule of $\vc'$ is an $i$-voter and $\vc'$ has at least one molecule.
A configuration $\vec{d}$ in $\calD'$ corresponding to $\vc$ may have some additional molecules of species $V_0$ or $V_1$.
The $i$-voters will eventually remove all molecules of species $V_{1-i}$ and will produce molecules of species $V_i$, but no molecules of species $V_{1-i}$.
Hence, eventually we reach a configuration $\vec{d}'$ with no molecules of species $V_{1-i}$ and at least one molecule of species $V_i$. We have that each configuration in $\post(\vec{d}')$ has this property.
In other words, $\vec{d}' \in \calO'_i$.
Hence $\calD'$ is a del-CRD.
\end{Proof}

\opt{sub}{
    \section{Proofs left out of main text} \label{sec:appendix-proofs}
    This section contains proofs of results that were omitted from the main text of this paper.

    \subsection{Closed upwards and semilinear sets} \label{sec:app:closedUpwardsSemlinear}
    \dicksonsLemma
    \relem{\ref{lem:closed_upw_semil}}{\lemClosedUpwardsSemilinear}
    \proofLemClosedUpwardsSemilinear
    \relem{\ref{lem:closed_upw_pre_post}}{\lemClosedUpwardsPrePost}
    \proofLemClosedUpwardsPrePost

    \subsection{Reactions can be assumed to have nonempty products} \label{sec:app:products-nonzero}
    This section proves the following lemma, which shows that we can assume without loss of generality that every reaction in a con-CRD has at least one product species.

    \relem{\ref{lem:no_empty_prod}}{\LemNoEmptyProd}
      \begin{Proof}
        \proofLemNoEmptyProd
      \end{Proof}

    Lemma~\ref{lem:no_empty_prod} is required to establish our assumption that $\pre(\vec{0})$ is semilinear, because
    \argumentPreZeroNotSemilinear

    \subsection{Dem-CRDs decide only semilinear sets}\label{sec:app:demCRDsemilinear}
    This section is devoted to proving Theorem~\ref{thm:dem-crd-semilinear}.

    \argumentDemCRDSemilinear
}

\end{document}